\documentstyle[chicago]{elsart} \newcommand{\Incl}[2]{}

\newcommand{\befig}{\begin{figure}[p]}\newcommand{\begtab}{\begin{table}}

\date{16 September 1997}

\newcommand{\X}{({\bf x})}

\newcommand{\x}{{\bf x}}
\newcommand{\y}{{\bf y}}
\newcommand{\w}{\omega}
\newcommand{\D}{{\bf \nabla}}

\begin{document}

\begin{frontmatter}

\title{Late states of incompressible 2D decaying vorticity fields}
\author{Enrico Segre\thanksref{corresp}}
\thanks[corresp]{Correspondence to Dr. Enrico Segre, 
    Dipartimento di Ingegneria Aeronautica e Spaziale, Politecnico, corso
    Duca degli Abruzzi 24, I-10129 Torino, Italy, tel.\ +39--11--564--6853,
    fax +39--11--564--6899, email {\tt segre@naima.polito.it}}
\address{Dipartimento di Ingegneria Aeronautica e Spaziale, Politecnico di
    Torino,Italy}
\and
\author{Shigeo Kida}
\address{Theory and Computer Simulation Center, National Institute for
   Fusion Science, Toki 509--52, Japan}

\begin{abstract}
Two-dimensional decaying turbulent flow is known to approach apparently
stable states after a long time evolution.  A few theories and models
have been so far proposed to account for this relaxation. In this paper,
we compare results of numerical experiments with the predictions of
these theories to assess their applicability. We study the long time
decay of initially multilevel vorticity fields on the periodic box, and
characterize the outcoming final states. Our final states do not match
the predictions of the theories; a broader variety of dipole profiles,
as well as nonstationary final states are found. The problem of the
robustness of the relaxational state with respect to variations of the
Reynolds number and different numerical resolution is addressed. The
observed configurations also do not necessarily possess the maximal
energy, in contrast to what is anticipated by some of the theories. We
are led to conclude that the mixing of the vorticity is generally not
ergodic, and that some metastable configurations may inhibit the
attainment of an equilibrium state.
\end{abstract}


\end{frontmatter}

\section{Introduction and motivations}

The decay of the vorticity field of a two-dimensional incompressible
fluid, which obeys to the unforced Navier-Stokes equation, shows several
interesting features. Stable large scale structures may form, organize
themselves on the scale of the accessible domain, and survive for a long
time before being ultimately damped by the dissipation. Occurrences of
such structures are observed in numerical and laboratory experiments, as
well as in planetary scale flows. We will call from here on such
configurations ``final''; by that we mean subjected only to viscous
decay and no more to filamentation and mixing. Provided that the
Reynolds number of the flow is large enough, the timescales for these
two processes can be very different.  We are
concerned with the appearance of final states emerging from
arbitrary initial conditions, with their robustness and their
predictability. Most of the numerical simulations presented in the
literature start from random initial conditions, whose Fourier spectra
decay variously in $k$, with uncorrelated phases. In such cases,
intermediate states with few isolated vortex cores are observed. In the
traditional scenario \shortcite{BePaSa88}, the progressive merging of likely
signed vortex cores is then observed. Several papers in the literature
deal with the merging process, either analyzing the evolution in time of
the vortex census, or modelling the process with simplified dynamics.
Less attention has been so far devoted to the actual characteristics of
the latest state of the field.

A few explanations for the occurrence of final states have been proposed
so far. We will refer specifically to those theories, which are
formulated in the physical space rather than in the Fourier one: the
minimum enstrophy assumption and the maximal entropy state statistical
theory \shortcite{MiWeCr92,RobSom91}. Some numerical
evidence has also been invoked to reconnect computed results with the
Joyce-Montgomery equation, which is originally derived for the mean
field limit of a system of point vortices \shortcite{MoShMa93}.

We consider several cases of decay of vorticity fields.
All cases start from initial conditions which should be generic and
appropriate for the application of the final state theories.
Our results contrast with the conclusion that an universal
state emerges out of the relaxation, without memory of the
dynamical path which led to it. In fact, we even see that the final
'state' may be unsteady: some fields attain a late time configuration
which is not stationary. Such late fields may move quasiperiodically and
steadily, being just slowly damped by the (arbitrary small) dissipation.

This paper is organized as follows: in section \ref{liter} we briefly
review the treatment of the problem of two-dimensional turbulent
decay, highlighting the different variational problems which are solved
to find the final state, and the functional  relations $\w(\psi)$
derived. In section \ref{esper} we describe the numerical experiments
performed.  Finally, in section \ref{Concl} we
comment on the open aspects which require further investigation.
A number of remarks which are related to the maximization of the energy and
its relevance for the final state are left for the Appendix.

\section{Review of the available theories} \label{liter}

Let us first recall the notation. As is well known, the equation of
motion for the vorticity $\w\X$ is written as
\begin{equation}
   \w_t\X = -{\bf v}\X\cdot\D\w\X + \nu\nabla^2\w\X
       = J(\w\X,\psi\X) + \nu\nabla^2\w\X\,.                  \label{eq:NS}
\end{equation}
In (\ref{eq:NS}), $\psi$ denotes the streamfunction, obtained from
$\w\X$ by means of the Green function $G$ of the Laplacian operator
 \begin{equation}
  \psi\X = \int G(\x',\x)\w(\x')d^2\x\,.
\end{equation}
The integral is carried on the fluid domain with the proper boundary
conditions, so that $\nabla^2\psi=-\w$. Introducing the
notation $\D^{\perp}= (\partial_y,-\partial_x)$, we may write
$\D^{\perp}\psi = {\bf v}$ and $J(\w,\psi)=\D\w \cdot\D^{\perp}\psi$.
The energy of the field is defined as
\begin{equation}
  E=\frac{1}{2} \int \psi\X\w\X d^2\x \, ,                     \label{eq:en}
\end{equation}
and the moments of the vorticity as
\begin{equation}
  Q_l=\frac{1}{2}\int \w^{l}\X d^2\x \,,
\end{equation}
with $l$ a positive integer. The quantity $Q_2(\equiv Q)$ is
traditionally called the enstrophy. Another global quantity which is
defined is the palinstrophy
\begin{equation}
  P=\frac{1}{2}\int \left[\D\w\X\right]^{2} d^2\x \;  .
\end{equation}
These quantities evolve in time according to:
\begin{eqnarray}
 E_t  = & - 2\nu Q \; , \qquad  \; Q_t = - 2\nu P \; ,
\nonumber \\
 Q_{l,t}  = & - l (l-1)\nu \int \w\X^{l-2}\left[\D\w\X\right]^{2} d^2\x \; .
\end{eqnarray}

$E$ and all $Q_l$ are constants of motion if $\nu=0$.
In the limit of vanishing viscosity,  $E$ is a constant of
motion, but $Q$ may not be, because $P$ can get larger inversely
proportional to $\nu$.

We refer to theories which predict the final state as the most probable
outcome of the decay. In a way or the other, all these models assume a
distinction between the fully detailed dynamics expressed by equation
(\ref{eq:NS}), and that of a reduced set of macroscopically observable
quantities. The evolution of the field is seen as a process in which the
initial information is lost in some way. As a deliberate simplification,
the final state is sought as the one which is fully described only by a
few macroscopical constraints, which are often called ``rugged
invariants''. Such theories treat the viscous, finite resolution
problem, as one in which some quantities are ``better'' conserved than
the rest, in place of the infinite set of the inviscid case. There is
indeed some ambiguity, and properly speaking the case $\nu=0$ is
different from the limit $\nu\rightarrow 0$, since in the former
infinitely steep gradients might form. Where possible, methods of
statistical mechanics are applied, and some extremum principle is
invoked. A comprehensive review of the various positions can be found
elsewhere (e.g.\ \shortciteNP{MiWeCr92}); here we recall them briefly, and
discuss their conclusions.

\subsection{Equilibrium Fourier spectra}  \label{eqSpectra}

A first group of theories is formulated in the Fourier space. The older
Krai\-chnan--Batchelor--Leith statistical theory \shortcite{Krai67} predicts
a Fourier spectrum $E(k)$ $\sim$ $k^{-3}$, relying on the assumption of
the locality of the interactions among the components. An improvement by
\shortciteN{Krai71}, based on the test-field-model closure
approximation, corrected this spectrum to $E(k)\sim k^{-3}
\left(\ln{{k}/{k_0}}\right)^{{1}\over{3}}$.
While some numerical simulations \shortcite{KiYaOh88,BMPS88} support
these spectra, quite different spectra have been observed by others
(see for example \shortciteNP{McWill84}) The reason of the discrepancy is not
clear, though the formation of stable structures may take a key role.

A later theory due to Kraichnan \shortcite{KraMo80,Carne82} proposes a
statistical mechanics for the energies of the Fourier components. An
ultaviolet cutoff in $k$ has to be enforced. Only $E$ and $Q$ are
assumed to be constants of motion, and are fixed as constraints. No
correlation is assumed between the phases of $\w(k)$, and no other
moment of the vorticity is conserved. This theory predicts a
statistical equilibrium spectrum
\begin{equation}
  E(k)=\frac{1}{\beta k^2 +\alpha}, \qquad Q(k)=\frac{k^2}{\beta k^2 +\alpha},
\end{equation}
with arbitrary constants $\alpha$ and $\beta$. The agreement of these
spectra with those coming from numerical simulations, and especially
with ours, is controversial.

\subsection{Point vortex systems}

A second line of reasoning considers the statistical properties of an
ensemble of point vortices. The rationale for connecting vorticity
fields with such systems is that a system of point vortices approximates
weakly, in the continuum limit, the Euler equation
\shortcite{CaLiMaPu92,EyiSpo93,CamOne93};  the full Navier-Stokes
equation can be emulated by point vortices which diffuse with an
additional Brownian motion \shortcite{chorin}. An entropy of the system
is introduced and maximized. In the mean field limit, a differential
equation is derived for the equilibrium configuration
\shortcite{MoJo74,Kida75,PoiLun76}:
\begin{equation}
 \w_0\X= -\nabla^2 \psi_0\X =
  c_1 \e^{-\beta \psi_0\X} - c_2 \e^{\beta \psi_0\X} \, .
                                                         \label{eq:JoMo}
\end{equation}
This provides us with a first example of an equation which relates
functionally $\w_0$ and $\psi_0$. As is known, the functional dependence
implies the stationarity of the motion, in the case of null dissipation.
In the special case of an equal number of opposite charged positive and
negative vortices, the Joyce-Montgomery equation reduces to the
sinh-Poisson equation
\begin{equation}
   \w_0\X=\lambda^2 \sinh(|\beta| \psi_0\X) \, .
\end{equation}
This equation has been furthermore studied referring to the inverse
scattering theory for the sin-Gordon equation \shortcite{TiChLe87}.
Solutions possessing simple scattering spectra can be constructed,
but no dynamical analysis, beyond a comparison of shapes, was done.

\shortciteN{MoShMa93} give an interpretation of
the problem which is somehow related. They propose a decomposition of
the vorticity field in four non-physical positive subfields.
An entropy of the form
\begin{equation}
  S=\int \w_i \ln \w_i d^2\x
\end{equation}
is then maximized individually for each subfield, with the proper
constraints on the total energy and vorticity. The remaining arbitrary
constants are fitted to the results of a single high resolution, high
Reynolds number Navier--Stokes numerical simulation. For that case, they
achieve a good fit of the $\psi(\w)$ scatterplot at late times.
For our purposes, it suffices to note that their final relation implies
\begin{equation}
  \w_0\X = c_1 \e^{-\beta\psi_0\X} - c_2 \e^{\beta\psi_0\X} + c_3 \, ,
\end{equation}
which is an elaboration of (\ref{eq:JoMo}), and can be assimilated to
equation (\ref{ompsiN}) in the case of 3 levels.

\subsection{Minimum enstrophy principle}

The identification of the final state as the one with lowest enstrophy
dates back to \shortciteN{BreHai76}. They argued that
while the energy can almost be conserved by a good numerical scheme, the
vorticity filamentates progressively and smoothes out. Arguments related
to the universality of the energy-enstrophy cascades, as in the
Kraichnan--Batchelor--Leith theory, would predict for instance a
behavior of $Q(t)\sim t^{-2}$ \shortcite{CaMWPo92,BarWar96}. The idea of a
faster decay of the enstrophy with respect to the energy is often
referred to as the ``selective decay hypothesis''. The final state is
consequently found variationally, by minimizing $Q$ with constrained $E$.
According to this hypothesis, axisymmetric vortex shapes on the infinite
plane can be calculated \shortcite{Leith84}. Time asymptotic estimates
for closed square box solutions are discussed by \shortciteN{vanGro88}.

In the case of doubly periodical boundary conditions, it is straightforward
to find a solution. Imposing
\begin{equation}
   \frac{\delta}{\delta\w}\frac{1}{2} \int (\w^2-\lambda\psi\w) d^2\x = 0\, ,
\end{equation}
we obtain
\begin{equation}
  \w_0(\psi_0)=\lambda\psi_0 \, ,                        \label{ompsiMinE}
\end{equation}
which on the periodic square admits solutions of the form
$\w_0\X=\sum\w_{{\bf k}} \e^{\i{\bf k}\cdot\x}$, with $|{\bf
k}|^2=\lambda$ equal to a squared integer. This implies that $Q=\lambda
E$. For a given energy the minimal enstrophy is then achieved for
$\lambda=1$, and the general solution becomes
$\w_0\X=\w_1\cos(x+a)+\w_2\cos(y+b)$, with $\w_1$, $\w_2$, $a$ and $b$
being arbitrary constants. Linear combinations of such sinusoidal
solutions with different $k$ (i.e. complete Fourier series) do {\em
not\/} satisfy the requirement. In the numerical experiments we find
final states of completely different forms. We remark however that this
principle was introduced for flows with additional ``topographic'' terms
in (\ref{eq:NS}), and we do not exclude that it may provide realistic
results in cases where these are dominant.

\subsection{Vortex censuses and punctuated dynamics}

In a number of papers, appeared around 1990, the late lowering of the
enstrophy is solely explained as a result of progressive vortex mergings
\shortcite{McWi90,MaStMaOu91,CaMWPo92}. These papers consider situations
with an intermediate time dynamics dominated by many well separated
vortex cores, which behave approximately like point vortices. The
subsequent evolution is schematized by a progressive collapse and merging
of these objects. Statistics of the number $N$ of cores in time, models
for the probability of merging are examined, and eventually lead to
different scalings of $Q(t)$. The final state is assumed to be a dipole,
and its properties are sought by scaling the relevant quantities down to
$N=2$. A popular model is the punctuated Hamiltonian system
\shortcite{CaMWPo91}, which is the traditional point vortex model fitted
with a nonconservative merging as vortices get close enough. Such
models can be further elaborated accounting for
extended cores \shortcite{RiPiBe95}.

\subsection{Maximum entropy theory}

The reasoning is based on the combinatorics of infinitesimal vorticity
patches, at a scale smaller than that which determines the energy of the
configuration. We follow the notation of \shortciteN{MiWeCr92}, rather
than the equivalent one of \shortciteN{RobSom91}. The theory is indeed
intended only for the Euler equation; \shortciteN{Wei93} proposes an
additional argument in order to include the viscosity in connection with
the underresolution, which seems incorrect. The theory mimics the
statistical mechanics of a many particle system. A given distribution of
(infinitesimally grained) vorticity is assimilated to a microstate; at
the macroscopical level, only a coarse averaged vorticity
$\overline{\w}\X$ can be observed. The fine scale structure is
remembered by introducing a local probability distribution
$n(\x,\sigma)d\sigma$ of vorticity, which says how large is the
probability of having a microscale vorticity $\sigma \leq \w\X <
\sigma+d\sigma$ at point $\x$. The macroscopic averaged vorticity is then
\begin{equation}
   \overline{\w}\X= \int \sigma n(\x,\sigma) d\sigma .
\end{equation}
The macrostate is the field of probability over the whole domain. Any
microscale distribution of vorticity, which looks on the macroscale like
that probability, is said to be a compatible microstate. The fluid is
expected to relax to the macrostate which can be achieved in the largest
number of ways, and thus is maximally probable. The theory relies on the
strong assumption, that the microscale mixing of vorticity is ergodic.
This means that the available vorticity is completely free to mix in any
possible (area preserving) way, so that only the probability determines
which outcome is likely to be observed. A statistical mechanics canonical
approach is undertaken. A free energy
\begin{equation}
    F(\{n\})= -S(\{n\}) - \beta E(\{n\}) + \sum_{l=0}^\infty  \mu_l Q_l.
\end{equation}
is maximized. Here $S(\{n\})$ is the entropy function
\begin{equation}
   S(\{n\})=-\int n(\x,\sigma) \ln n(\x,\sigma) d^2\x d\sigma \,,
\label{mixentro}
\end{equation}
the energy, expressed in terms of $n(\bf x,\sigma)$, is
\begin{equation}
E(\{n\})=
    \frac{1}{2}\int \sigma \sigma' n(\x,\sigma)
    n(\x',\sigma') G(\x',\x) d^2\x d^2\x' d\sigma d\sigma' \, ,
\end{equation}
and the constants of motion are
multiplied by appropriate Lagrange factors and added to $F(\{n\})$.

The constraints to be imposed are $E=\mathrm{constant}$, $\int n d\sigma
=1$ and $\int n d^2\x =g(\sigma)$. The function $g(\sigma)$ is the
global vorticity distribution, which should be invariant for microscale
inviscid flows. To implement the last constraint, the conservation of
all moments of the vorticity is instead required. It is assumed to be
sufficient that the infinite set of moments of the vorticity are conserved
without requiring the topological correctness of the flux. In other
words, it is assumed that the area preserving vorticity mappings are
dense (at least in the coarse average) in the topologically feasible
fluxes. Using $g(\sigma)$ instead of $Q_l$,
\begin{equation}
\sum_{l=0}^\infty \mu_l \int \overline{{\w}^l} d^2\x =
  \sum_{l=0}^\infty \mu_l \int \sigma^l  g(\sigma) d\sigma =
  \int \mu(\sigma) n(\x,\sigma) d\sigma  d^2\x \, .
\end{equation}
Functional derivation with respect to $n(\x,\sigma)$ and algebraic
manipulation lead to the system
\begin{eqnarray}
& & n_0(\x,\sigma) = \frac{\e^{-\beta(\sigma \psi_0\X-\mu(\sigma))}}
  {\int \e^{-\beta(\sigma \psi_0\X-\mu(\sigma))}d\sigma}, \nonumber \\
& & \w_0\X  =  \int \sigma n_0(\x,\sigma) d\sigma =
    -\nabla^2 \psi_0\X \,,                                  \label{n0cont}
\end{eqnarray}
which has to be solved in order to find the maximally probable
macrostate $n_0(\x,\sigma)$. In this procedure, the dependencies of
$\beta$ on $E$ and of $\mu(\sigma)$ on $g(\sigma)$ are left as implicit;
their actual form is supposed to be found only after solving
consistently the system. It is also assumed, but not proven, that
Lagrange multipliers can be determined for any physically accessible
values of the conserved quantities. The system (\ref{n0cont}) does not in
fact say very much. For $\beta<0$, it states that the probability of
having at $\x$ a vorticity of the same sign of $\psi_0\X$  grows with
$\sigma$, but is shaped by the weight factor $\exp[\beta\mu(\sigma)]$.
For $\beta>0$, the same applies to a vorticity opposite in sign to
$\psi_0\X$. The fact that $\beta<0$ in physical situations is inferred
in comparison with the case of a 2D Coulomb gas \shortcite{MiWeCr92}.

\shortciteN{RobSom92} also derive an evolution equation
for the approach to the maximum entropy state in this framework. Going
further on, \shortciteN{JuThTu96} propose a
way to include random forcing in the maximum entropy theory.

Particular forms of $\psi_0(\w_0)$ can be found only with additional
hypotheses. If, for instance, the vorticity takes only N different values
$(\w_1,\dots,\w_N)$, then
$n(\x,\sigma)$ will be everywhere a sum of delta functions.
The system (\ref{n0cont}) becomes
\begin{eqnarray}
n_0(\x,\sigma) & = & \frac{\sum_{i=1}^N
 \e^{-\beta[\w_i \psi_0\X-\mu(\w_i)]} \delta(\sigma-\w_i)}
  {\sum_{i=1}^N\e^{-\beta[\w_i \psi_0\X-\mu(\w_i)]}}
  = \sum_{i=1}^N n_i\X\delta(\sigma-\w_i) \,,                 \label{omegaNlev}
\\
\w_0\X & = &\frac{\sum_{i=1}^N \w_i\e^{-\beta \w_i
\psi_0\X}\e^{\beta\mu(\w_i)}}
  {\sum_{i=1}^N\e^{-\beta[\w_i \psi_0\X-\mu(\w_i)]}}
  =  \sum_{i=1}^N n_i\X\w_i  \,.                              \label{ompsiN}
\end{eqnarray}
The latter equation expresses a single valued relation $\w_0(\psi_0)$.
Inverted as $\psi_0(\w_0)$, it may eventually be multiple-branched.
Specifically, for two opposite levels $\w_1=-\w_2$, (\ref{ompsiN})
becomes
\begin{equation}
\w_0\X=\tanh\left(-\beta \w_1 \psi_0\X
  +\beta\frac{\mu(\w_1)-\mu(-\w_1)}{2}\right) \, .
\label{ompsiOppLev}
\end{equation}
The slope of the curve in the $(\w,\psi)$ plane is determined by
$\beta$, which depends on the energy; the position of the origin is
fixed by $\mu(\w_1)-\mu(-\w_1)$, which in turn can be expressed as a
function of $Q$ (all the higher moments of the vorticity are related to
$Q$ by the limitation to two levels).

For a two-level vorticity distribution it is even possible to
reconstruct $n(\x,\sigma)$ knowing $\overline{\w}\X$. This happens since
$n_1\X$ and $n_2\X$ are found from
\begin{equation}
 n_1\X+n_2\X=1 \,,  \qquad
\w_1 n_1\X+\w_2 n_2\X=\overline{\w}\X \,.
\end{equation}
The entropy $S$ can thus be evaluated directly in terms of
$\overline{\w}\X$:
\begin{eqnarray}
S & = & \int \frac{[\w_2-\overline{\w}\X]\ln[\w_2-\overline{\w}\X]
  + [\overline{\w}\X-\w_1]\ln[\overline{\w}\X-\w_1]}{\w_2-\w_1} d^2\x
\nonumber \samepage \\
  & & + \int \ln(\w_2-\w_1) d^2\x \, .
\end{eqnarray}

A variety of $\psi_0(\w_0)$ other than (\ref{ompsiOppLev}) can be
derived as well. Assuming {\em a priori\/} that
$\beta\mu(\sigma)=-|\sigma|/q$, that is, if the weight factor is
Poissonian \shortcite{Pas94},
\begin{equation}
\w_0\X = -2 \frac{\beta q^2\psi_0}{1-\beta^2 q^2\psi_0^2}.
\end{equation}
Assuming instead that $\beta\mu(\sigma)=-(\sigma/q)^2$, that is, if the
weight factor is Gaussian \shortcite[section VI D]{MiWeCr92},
\begin{equation}
 \w_0\X= - \beta q^2\psi_0   \,.
\end{equation}

A primary importance has been attributed to the `dressed vorticity
corollary' (DVC) \shortcite{MiWeCr92}. This corollary says that if one
``guesses'' $n_0$ from the macroscopical equilibrium state, that is if
one writes $n_0(\x,\sigma)=\delta(\sigma-\w_0\X)$, then a frozen
dynamics is obtained, which is the $\beta\rightarrow-\infty$ limit of
the true one. It is to remark that the ``dressed vorticity
distribution''
\begin{equation}
  g_d(\sigma)= \int \delta(\sigma-\w_0\X) d^2\x
\end{equation}
is, in general, different from $g(\sigma)$. This $g_d(\sigma)$ will be
approximated by the histogram of the vorticity distribution computed
from a numerical simulation. While $g(\sigma)$ would be conserved by an
inviscid dynamics, only $g_d(\sigma)$ may be constructed from the
computed field, and will vary in time. Only at the initial time, by
definition, $g(\sigma)=g_d(\sigma)$. A consequence of the DVC is that
the equilibrium field $\w_0$ is the one that has maximal energy among
all the fields with the  same distribution $g_d(\sigma)$.

\subsection{Applications of the maximum entropy theory}

Several recent papers compare direct numerical simulations with the
predictions of the maximum entropy theory. Even recognizing their value,
we think that their Ans\"atze are not justifiable in our cases, or that
their conclusions do not match our results. In detail,
\shortciteN{MiWeCr92} take into account a two-valued ($0$,$\w_0$)
and a three-valued ($0$,$\w_0$,$-\w_0$) initial condition onto a
circular corona with no slip boundaries. They solve the variational
problem by a Montecarlo dynamics and compare it with a direct numerical
simulation of the flow. \shortciteN{WhiTur94}
consider two equal circular vortex patches on a closed disk, with zero
ambient vorticity. They use an iterative solver for the constraint
equations, and compare the result with more extensive contour dynamics
simulations on the infinite plane (not on the disk). 
\shortciteN{SoStRo91} consider a shear layer, with periodical
boundary conditions in the $x$ direction and slip walls in $y$. Their
simulation starts from a two-level initial condition ($\w=0,a$), and
the nonlinear eigenvalue problem is solved accordingly. In a particular
limit, they are able to compute analytical solutions, which exhibit a
bifurcation in the parameter space. One of those branches corresponds to
solutions which break the symmetry in $x$, but are stationary (except for
a system of reference mean velocity). Their {\em a-posteriori\/} fit of
the  $\w(\psi)$ scatterplot is satisfactory only for one tract of
the complete curve. They also mention simulations of multiple shear
layers, saying that different local vortices are obtained, preventing
the system to achieve a steady state (``the system tends to achieve local
equilibria into vortices faster than the global equilibrium").
Additional simulations of an idealized jet are done by 
\shortciteN{ThSoJu94}; also in that case $x$-symmetry breaking
solutions are found from the maximum entropy theory, and a better
quantitative match is obtained. Symmetry breaking for periodic square,
periodic channel and box boundary conditions is further discussed along
those lines in a subsequent paper \shortcite{JuThSo95}.

In the maximum entropy setting, continuous symmetries generate additional terms
in the exponentials of the equations (\ref{n0cont}). If these terms
involve an explicit dependence on the coordinates, such as in the case
of the conserved angular momentum on the infinite plane, the soultion
of (\ref{n0cont}) could be non stationary \shortcite{RobSom91}. In the
periodic case, however, no continuous symmetry besides the translation exists,
and this possibility is prevented.

\shortciteN{ChaSom96} start assuming that $\w(\psi)$ is
linear, as it is in the minimum enstrophy context, and give analytical
maximum entropy solutions for rectangular and circular closed domains.
This is said to be justifiable in a particular `strong mixing' limit.
The solutions are always stationary, but admit monopole/multipole
bifurcations, which are thoroughly listed.

A true attempt to validate the maximum entropy rather than the minimum
enstrophy theory in an experiment is done by
\shortciteN{HuaDri94}. They consider a rather simple metaequilibrium profile
of a magnetized electron column, which obeys to the 2D Euler equation.
They conclude that the mixing cannot be assumed to be ergodic and that
the closest fit to the data is provided by the numerical minimum
enstrophy solution.

\section{Numerical experiments} \label{esper}

We performed a number of numerical experiments. To this extent we
integrated in time several vorticity fields, using a rather standard
protocol for the Navier--Stokes equation. We used a two-dimensional
$2/3$ dealiased pseudospectral code on the periodical square $(2\pi)^2$.
This choice, which fixes the Green function of the problem, was done for
easiness of implementation.  Nothing in the previously exposed theories
prevents us to use this particular choice of boundary conditions. The
Green function becomes $G({\bf k})=1/k^2$ in Fourier space, as known.
The (undealiased) resolution of the vorticity fields considered ranges
between $16^2$ and $512^2$. A small viscosity is introduced mainly for
numerical purposes, in order to prevent finite size effects, such as
high wavenumber pile-up. Viscosity is seen to be too small if the
energy, as seen from the $E(k)$ spectrum, accumulates at high $k$.
According to a generally accepted practice, we used hyperviscous
dissipative terms of the form $\nu_2\nabla^4\w$ or $\nu_3\nabla^6\w$ in
place of the ordinary viscous term. Even if drawbacks in the use of
hyperviscosity instead of normal viscosity have been recently discussed
\shortcite{Jimenez94}, we think that the choice is in practice uninfluent to
our simulation. We confronted runs in which the same initial conditions
were integrated with different forms of the dissipative term, observing
almost no influence on the resulting final configuration. The small
scale character of the hyperviscosity suppresses also some truncation
effects, such as Gibbs wiggles in the proximity of steep gradients of
the vorticity. Some dissipation is needed to `underresolve smoothly' the
smallest features which form during the evolution. The amount of
dissipation has to be grossly matched with the rate of creation of
smallest-scale structures, but does not have to be fine tuned. Time
marching is accomplished by a fourth order Runge-Kutta integrator, with
explicit treatment of the dissipative term. A Courant--Friedrichs--Levy
condition is employed to vary the time-step. The adoption of $\Delta t =
0.3 \Delta x / |{\bf v}_{\mathrm{max}}|$ appears to be accurate enough.

\begtab
\begin{tabular}{rrrrrrrrrr}
Fig. & res. & $\nu_3$ & $T_f$ & $E_0$ & $E_f$ & $Q_0$ & $Q_f$
   & $\diamond$ & $\equiv$ \\
\hline
\ref{fig:cover} & $400^2$ & $4\cdot10^{-13}$ & 449 &  $.01078$ &
  $.01072$ &  $.457$ & $.015$ & $+14\%$ & $-6.1\%$ \\
\hline
\ref{fig:zoopoli} & $256^2$ & $2\cdot10^{-11}$ & 745 & $.1052$ &
  $.1051$ & $.455$ & $.107$ & $+.26\%$ & $-10\%$ \\
  & $64^2$ & $4\cdot10^{-8}$ & 660 & .119 & .117 & .36 & .12
    & $+.17\%$ & $-11\%$ \\
  & $256^2$ & $2\cdot10^{-11}$ & 741 & .0138 & .0135 & .48 & .018
    & $+2.7\%$ & $-19\%$\\
  & $64^2$ & $4\cdot10^{-8}$ & 1047 & .034 & .032 & .319 & .037
    & $-.04\%$ & $-20\%$\\
  & $64^2$ & $1.9\cdot10^{-8}$ & 1291 & .0107 & .0082 & .431 & .012
    & $+2\%$ & $-30\%$\\
  & $256^2$ & $2\cdot10^{-11}$ & 343 & .16458 & .16454 & .29 & .20
    & $+.6\%$ & $-16\%$\\
\hline
\ref{fig:zoomovi} & $128^2$ & $2\cdot10^{-9}$ & 704 & .0178 &
  .0169 & .386 & .022
    & $+1.8\%$ & $-14\%$\\
  & $256^2$ & $2\cdot10^{-11}$ & 219 & .16417 & .16406 & .427 & .169
    & $-.5\%$ & $-2\%$\\
  & $256^2$ & $5\cdot10^{-12}$ & 267 & .1329 & .1328 & .466 & .144
    & $-1.2\%$ & $-4.8\%$\\
  & $64^2$ & $4\cdot10^{-8}$ & 182 & .3963 & .3961 & .445 & .433
    & $-12\%$ & $+1.4\%$\\
\hline
\ref{fig:4wavy} & $256^2$ & $2\cdot10^{-8}$ & 165 & .1639 & .1629 & .418
     & .184 & $-8\%$ & $+5.9\%$ \\
  & $256^2$ & $2\cdot10^{-9}$ & 175 & .1639 & .1634 & .418
     & .184 & $-9.4\%$ & $+4.3\%$ \\
  & $256^2$ & $2\cdot10^{-10}$ & 228 & .1639 & .1637 & .418
     & .186 & $-8.6\%$ & $+5.2\%$ \\
  & $256^2$ & $2\cdot10^{-11}$ & 293 & .1639 & .1638 & .418
     & .187 & $-11\%$ & $+3.5\%$ \\
\hline
\ref{fig:denti} & $512^2$ & $4\cdot10^{-12}$ & 340 & .1991
    & .1991 & .486 & .213 & $-4.5\%$ & $+3.7\%$ \\
  & $256^2$ & $2\cdot10^{-11}$ & 208 & .1991 & .1990 & .487
    & .213 & $-4.8\%$ & $+3.5\%$ \\
  & $64^2$ & $4\cdot10^{-8}$ & 367 & .199 & .198 & .486 &
    .202 & $-3.5\%$ & $+1.1\%$ \\
  & $32^2$ & $2\cdot10^{-7}$ & 419 & .199  & .196 & .467 &
    .199 & $-1.5\%$ & $-.2\%$ \\
  & $16^2$ & $1\cdot10^{-6}$ & 907 & .198 & .190 & .408 &
    .191 & $-1.3\%$ & $-3.9\%$ \\
\end{tabular}
\caption{Resolution, hyperviscous coefficient $\nu_3$,
  total integration time $T_f$, initial and final values of energy and
  enstrophy $E_0$, $E_f$, $Q_0$, $Q_f$ for the runs presented. The last
  two columns give the increase in energy of the ``prototype rearrangements''
  described in Appendix {\protect \ref{EmaxAlg}}, relative to $E_f$. $\diamond$
  stands for the `square dipole' and $\equiv$ for the `smooth stripe'.}
\end{table}

Most of our runs started from initial conditions consisting of
variously arranged constant patches of vorticity. In most cases we
simply distributed equal areas of vorticity equal to $\pm1$ on the
square. This already provided us with quite a variety of outcomes, and
allows direct connection with the various formulas of section
\ref{liter}. A pattern of this sort is the family of the `fuzzy
checkers'. By these we mean checkers with randomly perturbed edges. The
perturbation is done in a way that insures a constant zero mean
vorticity. A regular checker of vorticity is an unstable stationary
field (the velocity is orthogonal to the gradient of vorticity, which is
significantly different from zero only on the boundaries), while `fuzzy'
checker is immediately destabilized. Another possibility to generate
(usually) unstable initial conditions, is to consider random
arrangements of rectangular $\w=\pm1$ tiles upon the domain. A little
smoothing on the initial conditions is actually necessary to prevent the
edge ringing mentioned above, but, again, it does not seem critical for
the results. As remarked above, those initial conditions, even if a
little odd in appearance, are perfectly legitimate as test cases for the
relaxation models. The resolutions of the various runs presented,
together with the hyperviscous coefficient $\nu_3$, the total
integration time $T_f$, the initial and final values of energy and
enstrophy $E_0$, $E_f$, $Q_0$ and $Q_f$, are reported in Table 1.

\befig{\hspace{-0.6cm}  \vbox{\Incl{cover.ps}{14.6cm}}}
\caption{A typical case: decay of a two-level patched vorticity field
    into a stationary dipole. Plots of the vorticity field at various times.
    A few vorticity isolines are drawn for reference.}
    \label{fig:cover}
\end{figure}

A typical case is shown in figure \ref{fig:cover}. The patches initially
intertangle in a complicated way. The process generates a lot of
filamentary features, which, both because of the finite resolution and
the small-scale dissipation, are blurred out and disappear from the
landscape. At later times, the final shape becomes apparent. The process
of stretching and uniformization of the details continues, and a
`simpler' state stabilizes. This configuration undergoes no further
evolution, apart of a slow erosion of its contours, due to the
viscosity. Since the viscosity can in principle be made very small, this
state is long lived and can be named `final'. At this stage, the viscous
term of equation (\ref{eq:NS}) is some orders of magnitude smaller than
the vorticity (or the strain) field, everywhere on the domain. The
evolution of the velocity field is therefore still dominated by the
nonlinear terms, which balance to make the convective term null and the
vorticity configuration stationary. The timescale for viscous damping is
much longer than the characteristic turnover time.

\befig{\hspace{-0.8cm}  \vbox{\Incl{PU1x16n256_nu3_2e-11.ps}{14.6cm}
                  \Incl{PU64b4_mont_nu3_4e-8.ps}{14.6cm}
                  \Incl{PU16n256_nu3_2e-9_tv.ps}{14.6cm}
                  \Incl{PU64fz_2x2_nu3_4e-8_tv.ps}{14.6cm}
                  \Incl{PU64_8x8.ps}{14.6cm}
                  \Incl{PU256b8_Einc_nu3_2e-11.ps}{14.6cm}
                  }}
\caption{Arbitrary 2-level initial conditions which decay in final dipoles.
     Each row illustrates one case, with, in each panel: the initial
     and the late time vorticity field; the initial (dashed line) and final
     (full line) global vorticity distributions; the final
     scatterplot of ($\psi$,$\w$). All cases start from
     initial patches with $\w=\pm1$, except for the last one,
     where $\w=-1.5,+0.5$. The mean vorticity is always zero.}
     \label{fig:zoopoli}
\end{figure}

In the case of figure \ref{fig:cover} the final state is a dipole.
Something similar is also observed in other cases. We document a few of
them in figure \ref{fig:zoopoli}. The broadest vorticity contours in the
final state appear squared because of the bi-periodic boundary
conditions. This reflects the shape of the Green function
\shortcite{Glas74,Seyl76}. The maximum and the minimum of vorticity are
unique, and displaced each other by half of the box size in both
coordinates. The final state is stationary, and this agrees with the
traditional point of view, but not entirely. The profiles of the
positive and negative cores may differ. The scatterplot of $(\w,\psi)$
tends to a line, but does not match any of the expected functional
relations, such as (\ref{eq:JoMo}), (\ref{ompsiMinE}) or
(\ref{ompsiOppLev}). In all cases, the scatterplot of $(\w,\psi)$ is
seen to lie in the first and third quadrants. This is consistent with
any of the proposed $\w(\psi)$ relations, in which $\beta$ is negative.
In addition, the different cases mix differently the available
vorticity. If we plot the global distribution of the vorticity (third
panels of the graphs in figure \ref{fig:zoopoli}), we see that the final
histograms vary from case to case. The profile of the final vortex cores
is therefore peculiar to each decay. In all of them, however the final
vorticity has a single maximum and sigle minimum, and the dipole may be
thought as an arrangement of the avaliable vorticity around two centers.

\befig{\hspace{-0.6cm}  \vbox{\Incl{wavy.ps}{14.6cm}}}
\caption{A less typical case: decay of a two-level patched vorticity field
    into a wavy pattern.}
    \label{fig:wavy}
\end{figure}

The final dipole is not the only possibility, though. In figure
\ref{fig:wavy}, we see a different outcome. In this case a pattern of
two stripes is formed. The boundaries of these two stripes translate in
opposite directions varying slightly their curvature according to their
relative position. The state is therefore recurring. Interestingly, the
two zones cannot be explained simply by the absence of mixing of the
original patches: detailed analysis shows that each stripe contains a
fraction of the fluid of originally opposite vorticity. The mixing
inside a single stripe is complete, so that the vorticity is almost
constant and lower than that of the likely-colored patch in the initial
condition. The effectiveness of mixing within the stripe, but its
absence among the two stripes, is clearly evidenced when including
numerical passive markers into the flow (figures omitted).

\befig{\hspace{-0.8cm}  \vbox{\Incl{PU128fz3_nu3_2e-9.ps}{14.6cm}
                  \Incl{PU8n256_e_fz_nu3_2e-12.ps}{14.6cm}
                  \Incl{PU8n256_nu3_5e-12_tv.ps}{14.6cm}
                  \Incl{PU64fz_1x0_nu3_4e-8_tv.ps}{14.6cm}
                  }}
\caption{Arbitrary 2-level initial conditions which decay
      in nonstationary configurations. The format of presentation
      is the same as in Fig.\ \protect{\ref{fig:zoopoli}}.}
      \label{fig:zoomovi}
\end{figure}

In figure \ref{fig:zoomovi} we see other examples of fields which decay
in nonstationary final states. There is a variety of forms:
non-symmetric fat cores, which oscillate in time; wavy stripes with
central blobs that translate in one direction. Yet the last case shown
in figure \ref{fig:zoopoli} is not really a simple dipole; actually
three smaller and long lived vortices happen to form inside the main
core. They keep orbiting for long time and weaken very slowly because of
the viscosity, but are never able to merge. It is sometimes common, as
in the case of figure \ref{fig:wavy}, to get a complete uniformization
of the vorticity in different well-separated zones of the flow. A number
of patches of almost constant vorticity are thus formed, and their
arrangement prevents further mixing. In all these cases the scatterplot
of $(\w,\psi)$ never thins out, and, {\em a fortiori\/} cannot be fitted
by any of the relations given in section \ref{liter}. Nevertheless,
these configurations are long lived. Several features are still common
to both types of decays. The final states are ``dipoles'' in a broad
sense, in that they show two simply connected regions of vorticity above
and below the mean. The local maxima of $|\w|$ are also absolute maxima.
The small-scale structures are generated (and subsequently blurred)
quite early in the simulation. If we compute characteristic eddy
turnover times based on the sizes of the patches in the initial
conditions, we see that the palinstrophy (associated to the amount of
details) reaches its maximum within a few turnarounds. The entire
simulation is carried on typically for some tens of turnover times. The
loss of energy during the decay is always negligible. The enstrophy is
seen to decay significantly during the decay, until the final state is
formed; it then remains nearly constant. Most of the theories proposed
in section \ref{liter} take into account a lowering of $Q$, but fix it
in an unique way. We see that not even the initial $E$ {\em and\/} $Q$
are sufficient to determine the final state: as a counterexample, the
case of figure \ref{fig:wavy} and the one shown in the second row of
figure \ref{fig:zoomovi} possess initially nearly the same values of
$E$ and $Q$, and the same distribution of vorticity, but reach different
final states. Other counterexamples may be easily generated taking an
intermediate configuration and reversing the time integration. The
blurring of details is an irreversible process, and the backward
integration does not lead back to the initial condition. The forward and
backward final states have generally different properties.

For the initial conditions we have used, the final states appear to be
quite independent from the resolution and the (small) values and forms
of the viscosity. Generally, a higher spatial resolution just requires a
longer time to finish to smooth the details and to reach the final state.
To ground this affirmation we give some examples.

\befig{\hspace{-0.8cm}  \vbox{\Incl{PU256fz_128d_nu3_2e-8.ps}{14.6cm}
                  \Incl{PU256fz_128d_nu3_2e-9.ps}{14.6cm}
                  \Incl{PU256fz_128d_nu3_2e-10.ps}{14.6cm}
                  \Incl{PU256fz_128d_nu3_2e-11.ps}{14.6cm}
                  }}
\caption{Decay to the final wavy state, obtained with different
 hyperviscosities. The format of presentation is the same as in Fig.\
 \protect{\ref{fig:zoopoli}}. The value of $\nu_3$ is respectively 
 $2\cdot10^{-8}$, $2\cdot10^{-9}$, $2\cdot10^{-10}$, and $2\cdot10^{-11}$.}
    \label{fig:4wavy}
\end{figure}

First, we repeated the simulation shown in figure \ref{fig:wavy} with
four different values of the hyperviscosity. The appearance of the
field after long times is shown in figure \ref{fig:4wavy}. The same wavy
pattern, with uniform stripes of vorticity and undulated boundaries is
seen. Phase differences between the final panels are not meaningful,
since the field is plotted at different times. To demonstrate the
convergence and to provide one example of the evolution in time of
the characteristic quantities, we plot, in figure \ref{fig:4visco},
$E(t)$, $Q(t)$ and $P(t)$ for the different values of $\nu_3$. The
functions $E$ and $Q$ always stabilize on the final level. The loss in
$E$ is indeed negligible, less than $1\%$ in the most dissipated case.
The enstrophy instead decreases of roughly a factor two, but stabilizes
to a final value independent of the hyperviscosity. The palinstrophy
is seen to increase in time in the initial stages. This behavior, which
is also predicted by closure theories' estimates, is related to the
increase of small-scale details. The maximum value is achieved when such
details begin to be more rapidly numerically underresolved than
generated. This can be seen by looking at the field in physical space
(compare for instance the intermediate panels of figure \ref{fig:wavy}).
The curves are also seen to scale inversely to the hyperviscosity.
This behavior, apart from different values and different relaxation times,
is also found in all the other decays, both the stationary and
nonstationary ones. It is also observed when normal viscosity or
second viscosity are employed.

\befig{\Incl{4visco3.ps}{14.0cm}}
\caption{Energy, enstrophy and palinstrophy decay
     for different values of the hyperviscosity.}
\label{fig:4visco}
\end{figure}

Secondly, in order to show that the final state found is not an artifact
of the finite resolution, we display the integration of another
arbitrary initial condition at different spatial resolutions (figure
\ref{fig:denti}). The hyperviscosity coefficients are adjusted case by
case according to the highest wavenumber retained. We see that unless
the resolution is extremely poor ($16^2$), a final state with given
characteristics is obtained. In this case, the final state is
nonstationary, the two elongated cores translate in one direction, while
both background stripes convect in the opposite one.

\befig{\hspace{-0.8cm} \vbox{\Incl{PU512_dente_nu3_4e-12.ps}{14.6cm}
                  \Incl{PU256_dente_nu3_2e-11.ps}{14.6cm}
                  \Incl{PU64_dente_nu3_4e-8_tv.ps}{14.6cm}
                  \Incl{PU32_dente_nu3_2e-7.ps}{14.6cm}
                  \Incl{PU16_dente_nu3_1e-6.ps}{14.6cm}}}
\caption{Effect of lowering the resolution of the simulation. The same initial
   condition is integrated at $512^2$, $256^2$, $64^2$, $32^2$, $16^2$, and
   hyperviscosity rescaled appropriately. The format
   of presentation is the same as in Fig.\ \protect{\ref{fig:zoopoli}}.}
   \label{fig:denti}
\end{figure}

We note, passing by, that the Fourier spectrum of these final states is
always decreasing in $k$. We are not however concerned with possible
laws to fit their slopes, and we do not report them here. In general,
the final state is neither a pure $|{\bf k}|=1$ state (as predicted by
the minimum enstrophy principle), nor a spectrum introduced in section
\ref{eqSpectra}. Most of the energy of the final field is contained in
the first modes, but this fact itself does not explain the formation of
a huge variety of profiles.

\section{Conclusions and perspectives} \label{Concl}

We have considered several examples of relaxations of two-dimensional
vorticity fields, and compared them with the theories which have been so
far proposed in literature. We have brought some numerical evidence that
contrasts with the existing final states theories. The different
theories do not agree in their predictions, as can be seen in the
different $\w_0(\psi_0)$ which they propose. We have shown the existence
of legitimate final states which are not included in the accepted point
of view. Nonstationary ones are among them. The remark that stable,
nonstationary configurations may indeed form from unstable ones is
indeed not new. Stable tri- and quadrupole satellite systems, which
result from the decay of unstable vortex blobs, have already been
investigated analytically and numerically
\shortcite{CaFlPo89,Carton92,MorCar94,CarLeg94}. Beautiful experimental
evidence is also provided \shortcite{HeKlWi91,FlovHe96}. In all these cases
$(\w,\psi)$ is derived from the experiment, and is seen to evolve from a
non-monotonous to a branched monotonous curve. On the other hand, also
the non-uniqueness of stable dipolar solutions on the periodic box has
been stressed \shortcite{RaHeSc96}. Numerical experiments shown there display
$(\w,\psi)$ scatterplots which are tentatively fitted by polynomial
relations. For this reason we have not tried to fit one or the other of
the free parameters of the reported models versus our numerical results:
the possibility of a good fit appears more accidental than general. In
our opinion many of the underlying assumptions of these theories are not
always granted, and call therefore for a more careful treatment. For
instance, we think that the final state cannot be predicted by a 'final
state theory' which ignores completely the dynamical path underwent by
the relaxing system. This is demonstrated by the fact that
configurations with the same two initial vorticity levels decay to
different final states. The quest for appropriate statistical
descriptors, which are preserved during the evolution, is still open.
For instance, it is not known how the initial macrovorticity
distribution relaxes to the final one. It would be appealing to
parametrize this change by a few dynamical quantities, such as the
decrease of $Q$. We have supposed that some of the shapes of the final
states can be understood as rearrangements of the vorticity which is
available at late times. This seems to apply at least to the final
stationary states, for which the requirement of energy maximization
appears as sufficient. The instability of the intermediate vorticity
configurations is always such that the system is led to mix its
vorticity and drawn toward the final state. In the other cases we may
conjecture that, for some stability reason, the available vorticity is
unable to collapse onto a stationary configuration. In many of the
previous works, initial Gaussian fields with random spectra were
considered. These fields show generally a smaller population of absolute
higher vorticity (i.e., $g_d(\sigma)$ with long tails). It is often
observed (but not always) that a lot of
concentrated vorticity patches are formed in the early times of
evolution and they survive relatively long times 
(see e.g. \shortciteNP{BMPS88,Kida85}). Such intermediate
states, composed of several isolated cores, which have so often been
described in literature, could be understood as metastable preliminary
local rearrangements of the vorticity.

In other cases, however, the attainment of stationary states seems
prevented. Possibly, the system is trapped in a metastable state, which
is not stationary (in the cases encountered in this work, periodical in
time). The stationary state could be missed just because the
dynamics of the vorticity is by itself insufficient to trigger a
catastrophic mixing process, which alone could alter further the
vorticity distribution, and lead the system to its very final state. It
would then be very interesting to find the requirements for this
metastability. An implication of such metastabilities would be that some
kind of external forcing might trigger the transition to the stabler
states and some perhaps not.

\section*{Acknowledgements}

The motivation of this work originates from discussions with
Professors R.A. Pasmanter and K. Ohkitani while one of the authors (S.K.)
visited the Koninklijk Nederlands Meteorologisch Institut, under the
support provided by the Dutch Science Foundation NWO, Priority Programme
on Nonlinear Dynamical System. The major part of this work was then
carried on at the Research Institute for Mathematical Sciences of the
University of Kyoto. E.S. was supported  during that period by the
Japanisch-Deutsches Zentrum Berlin Sonderaustausch Program. We would like to
express our gratitude to Prof.\ R.\ Pasmanter, Prof.\ K.\ Ohkitani, Dr.
G.\ Boffetta and Dr. A.\ Celani for their comments and fruitful
discussion.

\renewcommand\thesection{Appendix:}
\renewcommand\thesubsection{A.\arabic{subsection}}

\section{Energy maximization and stability} \label{MaxE}

\renewcommand\thesection{A}

All the theories proposed in section \ref{liter} identify the final
state as the one which solves a certain extremum problem. In the case of
the minimum enstrophy principle, one may think to rescale the vorticity,
and to rephrase the problem as to the search for the maximal energy,
given a fixed enstrophy. In the setting of the minimum entropy, the DVC
instead tells us that the final state is actually also a state of
maximal energy (with fixed $g_d$). It is therefore worth to add some
remarks on energy maximization.

We are not able to determine in how many equally energetic ways
the same amount vorticity can be distributed over a domain. We might ask
if equally energetic states which are not dynamically related, i.e.,
which do not evolve one into the other, because they are topologically
different, could converge to the same final state. We cannot answer to
those questions, but we can add a few considerations which at least rule
out some possibilities. These provide necessary
requirements, but they do not determine the configuration of vorticity nor
the flow.

\subsection{Infinitesimal deformations} \label{InfDef}

Let us first consider a generic area preserving infinitesimal
deformation on a two-dimensional domain. The deformation may be written
as $\delta\x=\D^{\perp}\delta\phi\X$, where $\phi$ is an arbitrary
field. It is easily seen that such transformations preserve all the
moments of the vorticity:
\begin{eqnarray}
\delta\int \w^l d^2\x &=&
 l \int J(\w,\delta\phi)\w^{l-1} d^2\x = \nonumber \\
 &=& -l \int\delta\phi\left[(\w_x\w^{l-1})_y-(\w_y\w^{l-1})_x\right] d^2\x
= 0 ,
\end{eqnarray}
Integration by parts is carried out since the
boundary terms vanish for any boundary condition.
The variation of the energy with respect to such
deformations is
\begin{eqnarray}
\delta E &=& \delta{\textstyle\frac{1}{2}}
 \int \w\X\w(\x') G(\x',\x) d^2\x d^2\x' =
 \int \delta\w\X \w(\x') G(\x',\x) d^2\x d ^2\x' = \nonumber \\
  &=& -\int J\left(\delta\phi\X,\w\X\right) \w(\x') G(\x',\x) d^2\x d^2\x'=
 \nonumber \\
  &=& \int \delta\phi\X J\left(\w\X,G(\x',\x)\right) \w(\x') d^2\x d^2\x' \, .
\end{eqnarray}
Therefore the extremal energy states satisfy
\begin{equation}
\frac{\delta E}{\delta \phi}= \int J\left(\w\X,G(\x',\x)\right) \w(\x')d^2\x' =
 J\left(\w\X,\psi\X\right) = 0.
\end{equation}
Thus if the dissipation is absent, the stationary states
extremize the energy with respect to infinitesimal incompressible
deformations. These states are ``local'' extrema.

\subsection{Pointwise mixing exchanges} \label{MixExc}

We can write a transformation which transfers some of the vorticity
present in the neighborhood of $\x_1$ to a similar neighborhood of
$\x_2$ and vice versa. To this extent we employ a shape function
$\delta_\epsilon\X$, which is equal to ${1}$ in a neighborhood of
radius $\epsilon$ of $\x=0$ (modulo the boundary conditions) and zero
everywhere else. The vorticity field is transformed according to
\begin{eqnarray}
\w\X \rightarrow \w\X
 &+& r \left[\w(\x-\x_1+\x_2) -\w\X\right] \delta_\epsilon(\x-\x_1)
\nonumber \\
 &+& r \left[\w(\x+\x_1-\x_2) -\w\X\right] \delta_\epsilon(\x-\x_2) \, .
\end{eqnarray}
The parameter $r$ may vary from $0$ to $1$; in the latter case, an
exchange between the vorticity in $\x_1$ with that in $\x_2$ is
realized. The energy of the transformed field is computed by substituting
the above expression in equation (\ref{eq:en}). When consider the limit
$\epsilon\rightarrow 0$, the integrals are approximated by the mean
value of the integrand times the area of the integration region,
provided that $\w$ is continuous. The terms multiplied by $r$ come of
order $O(\epsilon^2)$, while the ones with $r^2$ remain $O(\epsilon^4)$.
The singularity in the Green function poses no problems, since for small
arguments $G(\x,\y)\sim -\ln|\x-\y|$, which is integrable. Therefore
\begin{equation}
  E \rightarrow E + r \pi \epsilon^2 [\psi(\x_2)-\psi(\x_1)]
[\w(\x_1)-\w(\x_2)]
   + r^2 O(\epsilon^4) \, .
\label{maxe:scambio}
\end{equation}
This gives us the infinitesimal change in energy due to a transformation
which modifies infinitesimally, but not continuously, the field. The fact that
this transformation is not dynamical is not of our concern, as far as we
look for properties of the configurations and not for their evolution.

If we refer to the maximal energy state, its energy shall decrease
whichever the couple of points ${\bf x_1}$ and $\x_2$ and the value of
$r$. Therefore, $\w(\x_2)>\w(\x_1)$ implies $\psi(\x_2)>\psi(\x_1)$ and
vice versa. The most important consequence of it is that 
$\psi(\w)$ must be single valued and monotonous. This is seen by
considering the plot of $\w\X$ versus $\psi\X$. The monotonicity is
implied by the fact that data points on the $(\w,\psi)$ plane can be
labelled by $\x$. If the relation $\w(\psi)$ was non monotonous or more
than single valued, then it would be possible to find points at which
$\w\X\le\w(\x_1)$ and $\psi\X\ge\psi(\x_1)$,
contrary to the hypothesis. As a further consequence, the absolute maximal
energy state is also stationary, because $\w\X$ and $\psi\X$ are
functionally related.

A particular case is achieved when the points are infinitesimally
close and displaced in the direction ${\bf d}$: in this case
(\ref{maxe:scambio}) is read as $({\bf d}\cdot\D\psi)({\bf
d}\cdot\D\w)>0$ at any point and along any direction.

Related to this, but derived with a different machinery, is the
Rayleigh--Arnold criterion \shortcite{HoMaRaWe,MISh87,CarVal90}. It states
that among all the isovortical fields the maximal energy arrangement is
unique and stationary. Furthermore, if ${d \psi_0(\w_0)}/{d \w_0}$, or
equivalently $-{\D\psi_0\cdot\D\w_0}/{|\D\w_0|^2}$ are strictly positive
and limited, then the field is also nonlinearly stable. We also note
that there may exist also other stable states, in particular those
obtained by means of some symmetry transformation, which is specified by
invariants inexpressible as functionals of the vorticity. Analytical
stability criteria other than Arnold's one, that refer to local features
in ambient flows (e.g.\ \shortciteNP{Nycand95}), do not appear to be of direct
use for assessing global stability.

\subsection{Energy maximization algorithms} \label{EmaxAlg}

We have seen in \ref{InfDef} and \ref{MixExc} properties of
`energy-\-maximizing arrangements of the available vorticity'. By
`arrangement', we mean another field, which has exactly the same
fractional area for each level of vorticity (i.e. the same vorticity
histogram), but a different configuration. The preceding facts tell us
that these rearrangements are unique, stationary and stable. We are not,
however, guaranteed that {\em any\/} initial field relaxes into such an
energy maximizing configuration. Not only it is {\em a priori\/}
undetermined how the histogram of macrovorticity evolves in time, but
also not all fields are seen to approach a stationary final state. We
want therefore to characterize the maximal energy state for a given
vorticity population, and its relation with the observed dynamical final
state.

One possibility would be provided by the Carnevale, Vallis and Young
pseudodynamics \shortcite{VaCaYo89,CarVal90}. They proposed a numerical
procedure to deform a vorticity field by increasing its energy and  
conserving its enstrophy. The procedure does not explain the geometrical
properties of the extremal states, which can only be said to be local,
and not absolute maxima of the energy. Lacking analytical procedures, we
rather used a simpler alternative combinatorial approach. Starting from
a given configuration of vorticity, two square tiles are randomly
exchanged and the new energy is computed. The new configuration is kept
if the energy increases, discarded otherwise, and the process repeated.
\shortciteN{MiWeCr92} also use a more sophisticate
numerical procedure of this kind. Such approach is easy to implement,
but very slow in convergence when the number of tiles is significant.
With the aid of this tool, we observed that the highest energy
configuration achieved by permutation of equal area tiles of vorticity
$\pm\w_1$ on the periodic square, is a subdivision in two stripes
parallel to the sides. When intermediate levels of vorticity are
present, as in our runs, where patches' edges are significantly smoothed,
the most energetical of these permutations of tiles has sometime still a
striped shape, sometimes a central cored appearance.

To cope with high resolution rearrangements, we chose to limit ourselves
to compare energies of only a few selected prototypical configurations.
To produce them we sorted the discrete tiles of the field according to
value of their vorticity, and deployed them upon the domain following
the same order of a sample configuration. In this way we obtain a
rearrangement with the same set of contour lines of a given prototype.
We can then check which, among few prototype rearrangements, increases
the current energy.  One of the test profiles we chose is the ``square
dipole'', which is modeled on
$G\X-G\left(\x+\left(\pi,\pi\right)\right)$; its contour lines look
similar to those of various dipolar final states shown in Fig.\
\ref{fig:zoopoli}. Another one is the parallel stripe, which is the
maximal arrangement for a two-level population. If we compute the
energies of the rearrangements of the final states obtained in section
\ref{esper}, we basically see that for the stationary final states the
highest energy is achieved for the `square dipole' rearrangement, with
minimal increase of energy. For the nonstationary final states, such as
the wavy ones, often the stripe rearrangement results more energetical.
The actual values of the energies are reported in Table 1. The choice of
the maximal profile is basically a function of the available vorticity.
Still, the energy of the nonstationary states is often significantly
increased by the procedure. This indicates that the final state, in
those cases, is far from the highest energy configuration for the given
$g_d$. Conversely, the final state may be dynamically inaccessible
from the rearrangement: such is the case for the wavy configurations, if
we maintain that the energy is almost conserved. In addition, the wavy
cannot be seen as a perturbed stripe, since the maximal energy state
cannot be unstable to transversal perturbations, thanks to the Arnold
stability criterion.

\befig{\hbox{\Incl{schema.ps}{5.9cm} \Incl{surfe.ps}{7.8cm}} }
\caption{Energy of a family of wavy patterns of vorticity equal to $\pm1$,
         as a function of $A$ and $\Delta \phi$. The region of the
         $(A,\Delta \phi)$ plane where the two stripes would interfere is
          drawn cross-hatched.}
\label{fig:Ewavy}
\end{figure}

We are not able to provide a full analytical treatment of the family of
the recurring wavy solutions here, but we would like to add a small piece of
numerical evidence.  We plot the energy corresponding to all two level
striped fields with sinusoidal undulations, with varying amplitudes and
phase shift (figure \ref{fig:Ewavy}). We see that the energy has to
decrease significantly with increasing amplitude in order to allow for
undulations of the boundaries. For a range of energies below the
highest, energy isolines span the entire interval
$0\le\Delta\phi\le2\pi$, with the amplitude depending on the phase
shift. A recurring evolution of the field, with pulsating amplitudes, is
therefore allowed. Without attempting to make a precise fit to our final
states, we claim that the picture is appropriate. It is clear from the
figure that for amplitudes of the undulation higher than $A\sim {1\over
4}\pi$, but still lower than the geometrical limit $A={1\over 2}\pi$,
the range $0\le\Delta\phi\le2\pi$ is no more entirely accessible at
constant $E$ (the separatrix is drawn in bold). In fact, numerical tests
with initial conditions of this kind exhibit stability below an
amplitude threshold, and a catastrophic mixing above.

\bibliography{psichojitensha9}
\bibliographystyle{chicago}

\end{document}